\documentclass[a4papper,oneside,fleqn,notitlepage,9pt,twocolumn]{amsart}
\usepackage[a4paper]{geometry}

\usepackage[cm]{fullpage}
\usepackage[dvips]{graphicx}
\usepackage{xspace}
\usepackage{amsmath,mathtools}
\usepackage[]{kpfonts}
\usepackage{xfrac}
\usepackage{urwchancal}
\usepackage{wrapfig}
\usepackage[framemethod=TikZ]{mdframed}

\renewcommand{\eqref}[1]{Eq.\ref{#1}}
\renewcommand{\vec}[1]{{\bf{#1}}}
\renewcommand{\wp}{\mathpzc{p}}
\renewcommand{\sinh}{\mathrm{sh}}
\renewcommand{\cosh}{\mathrm{ch}}
\renewcommand{\tanh}{\mathrm{th}}
\renewcommand{\ell}{\mathpzc{l}}

\DeclareMathAlphabet{\mathpzc}{OT1}{pzc}{m}{it}

\newcommand{\abs}[1]{\left|#1\right|}

\newcommand{\br}[1]{\left(#1\right)}
\newcommand{\cm}{{\it CM}\xspace}
\newcommand{\dk}{\dot{k}}

\newcommand{\dx}{\dot{x}}
\newcommand{\etasign}{}
\newcommand{\elm}{\mathpzc{m}}
\newcommand{\eqdef}{\!:=\!}
\newcommand{\eqrefs}[2]{Eqs.\ref{#1},\ref{#2}}
\newcommand{\gramm}[2]{\mathrm{Gram}\left(#1,#2\right)}
\newcommand{\Kpi}{\pi}
\newcommand{\mcal}[1]{\mathpzc{#1}}

\newcommand{\PB}[2]{\left\{#1,#2\right\}}
\newcommand{\perpp}{{\scriptscriptstyle\perp}}

\newcommand{\sgn}[1]{\mathrm{sgn}\left({#1}\right)}
\newcommand{\secref}[1]{Sec.\ref{#1}}
\newcommand{\seq}{\equiv}
\newcommand{\slot}{\square}
\newcommand{\SP}[2]{\langle#1\!#2\rangle}

\newcommand{\ts}{\textstyle}

\newcommand{\udp}[1]{\mathfrak{d}_{\tau}\!#1}
\newcommand{\uddp}[1]{\mathfrak{d}^{\scriptscriptstyle2}_{\tau}\!#1}
\newcommand{\udddp}[1]{\mathfrak{d}^{\scriptscriptstyle3}_{\tau}\!#1}
\newcommand{\udnp}[2]{\mathfrak{d}^{\scriptscriptstyle{#1}}_{\tau}\!#2}
\newcommand{\weq}{\simeq}

\newcommand{\wq}{\mathpzc{q}}
\newcommand{\wdq}{\dot{\mathpzc{q}}}

\begin{document}

\title{Relativistic Ideal Clock}
\author{{\L}ukasz Bratek\\{}\\
   Institute of Nuclear Physics,\\ Polish Academy of Sciences,\\
Radzikowskego 152, PL-31342 Krak\'{o}w, Poland\\
   \texttt{Lukasz.Bratek@ifj.edu.pl}}
\date{\today}

\maketitle

\begin{abstract}
Two particularly simple ideal clocks exhibiting intrinsic circular motion with the speed of light and opposite spin alignment are described. The clocks are singled out by singularities of an inverse Legendre
transformation for relativistic rotators of which mass and spin are fixed parameters. Such clocks work always the same way, no matter how they move.  When subject to high accelerations or falling in strong gra\-vitational fields of black holes, the clocks could be used to test the clock  hypothesis.
\end{abstract}  


\bigskip

An ideal clock is a mathematical abstraction of 
a nearly perfect material clocking mechanism.
The clock hypothesis asserts that an ideal clock measures its proper time. This means that the 
number of consecutive cycles registered by the 
clock increases steadily with the affine parameter of the worldline of the  clock's center of mass (\cm). 
On the dimensional grounds, we may expect that the hypothesis could be violated for extreme accelerations of order $c\omega$ (e.g. \!$\tfrac{mc^3}{\hbar}\!\sim \!10^{^{_{29}}}\!\,\!\tfrac{\mathrm{m}}{\mathrm{s}^2}$ for the electron).

A recent result \cite{bib:QFT} suggests that quantum field-theoretical realizations of extended clocks (experiencing the Unruh effect) do not measure their proper times. But the clock hypothesis refers to classical concepts of the relativity theory (e.g. a single worldline), and as such should be first of all tested within the same conceptual framework.  A candidate clock should be a relatively simple mathematical device so as to minimize the influence of external disturbances on its structure.  If some fundamental limitation were to concern such a clocking standard, the more it would concern more complicated clocks.

A mathematical clock can be devised by an analogy with a quantum particle such as Dirac electron. The intrinsic clock of such particle cannot be impaired -- the phase of its wave function oscillates in the rest frame with a fixed frequency determined by only
the fundamental constants of nature. But quantum phase is not observable, it would be useless as a clock. Something similar happens with the basic classical analogue of a quantum particle -- a structureless material point. The action functional of the material point (to some extent related to a quantum phase) increases linearly with the affine parameter of its worldline. But classical observables are re\-pa\-ra\-me\-te\-ri\-za\-tion invariant, and do not distinguish any particular time variable. In order to play the role of  an ideal clock, the material point must be endowed with an additional structure repeatedly changing with the proper time, e.g. connected with some sort of intrinsic rotation.  Additionally, for the clock to resemble a quantum particle with its invariable structure as much as possible, it may be required that the clock's mass and magnitude of its spin should have fixed numerical values.       

\section{\label{sec:mech} A relativistic clocking mechanism}

In the relativity theory, a rotation can be described  as a continuous action of an elliptic homography mapping points on a complex plane at one instant to those at another instant and leaving fixed  
a pair of points: $\kappa_+$ and $\kappa_-$. It is natural to identify $\kappa_{\pm}$
with stereographic images $\mcal{Z}(k_{\pm})$ of a pair of null vectors preserved in free motion of any isolated massive system with spin: $$k_{\pm}\equiv\frac{p}{\sqrt{\SP{p}{p}}}\pm\frac{w}{\sqrt{-\SP{w}{w}}}.$$ Here, $p$ is the momentum vector 
and $w$ is the Ma\-this\-son  pseudovector 
 (customarily ascribed to Pauli and Luba\'{n}ski, see \cite{bib:trautman}). 
Given a pair $k_{\pm}$, the homography is set by specifying the motion 
of a single point $\kappa\equiv\mcal{Z}(k)$ -- a stereographic image 
of a null vector $k$ -- about an invariant circle of that homography. 
The $\kappa$ is invariant under a local scaling $\delta{k} =\epsilon k$, so should be the Lagrangian. 
It is thus necessary that
$\SP{k}{\Kpi}= 0$ identically (with $\Kpi$ being the momentum conjugate to $k$), since otherwise the variation 
$\delta{L} = \epsilon({ k\partial_k{L} +
\SP{\dk}{\Kpi}})+ \SP{k}{\Kpi}\dot{\epsilon}$ would not be vanishing for arbitrary $\epsilon$. Accordingly, there are two structure constraints:
\begin{equation}\label{eq:fundcond2}
\Psi_3:\quad\SP{k}{\Kpi}\weq0,
\qquad \Psi_4:\quad\SP{k}{k}\weq0.
\end{equation}

\begin{wrapfigure}[7]{R}[0\textwidth]{0.185\textwidth}
\vspace{-1.3\baselineskip}
\includegraphics[width=0.185\textwidth,
, clip
]{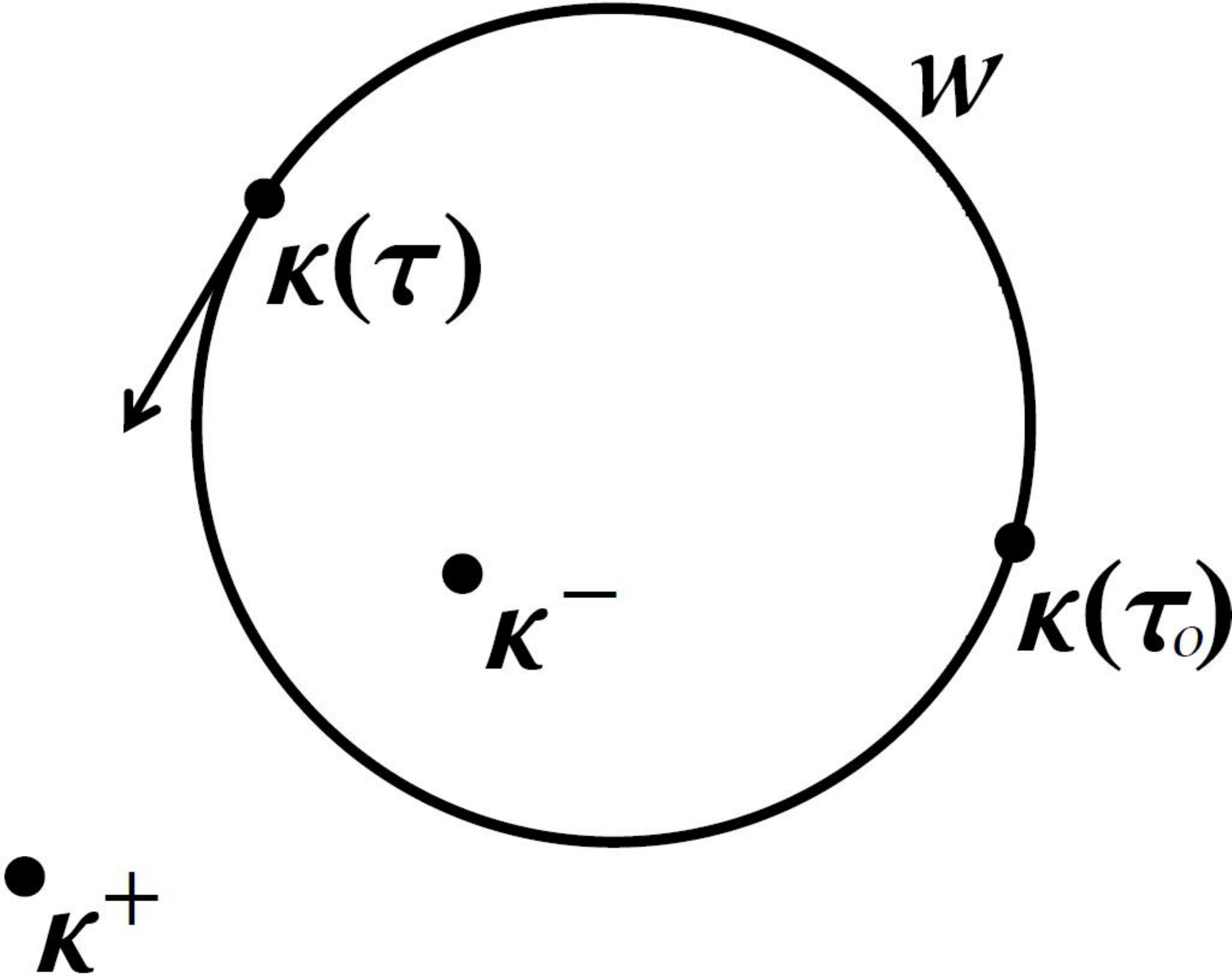}
\end{wrapfigure} 

\noindent
Now, it can be deduced  what the invariant circle must be.
From
$w\propto \ast (p\wedge k \wedge \Kpi)$ it follows
that $\SP{k}{w}=0$. This means that
the image point of $k$ moves about the image circle of $w$. As so, $k$ may be thought
of as representing the clock's pointer and $w$ as representing the clock's dial (see figure).
In free motion in Minkowski spacetime, 
 the vectors $k_{\pm}$
are parallel transported. Then a Lorentz invariant phase can be assigned to $\kappa$ between instants $\tau_o$ and $\tau$ through: 
\begin{equation}\label{eq:phase}
\phi
(\tau,\tau_o)
=i\,\mathrm{Ln}\,{
\br{\frac{\kappa(\tau)-\kappa_{+}}{\kappa(\tau)-\kappa_{-}}\cdot
\frac{\kappa(\tau_o)-\kappa_{-}}{\kappa(\tau_o)-\kappa_{+}} }}\,.\end{equation}
The phase $\phi$ is a real number. In free motion of the clock, a rotation through $\phi=2\pi$
represents a single full clocking cycle.\footnote{A massless system  ($\SP{p}{p}=0$) would be structurally  
something different, because for a 
parabolic homography (preserving a null direction $p$ and an orthogonal to it spatial direction $w$) the analogous phase is not a Lorentz scalar.} 

\section{\label{sec:hamiltonian}Dynamical requirements  and the  Hamiltonian }

The intuition derived from the theory of Eulerian rigid bodies suggests that the the above clock will be insensitive to external influences if both its mass and size is fixed. This requirement can be fulfilled in an invariant way by imposing constraints on the Casimir invariants of the Poincar\'{e} group: 
\begin{equation}\label{eq:fundcond}\Psi_1:\quad\SP{p}{p}-\elm^2\weq0,
\qquad\Psi_2:\quad\SP{w}{w}+\frac{1}{4}\elm^4\ell^2\weq0.\end{equation} Here, 
constants $\elm$ and $\ell$ are fixed parameters with the dimension of mass and that of length.
These constraints should be regarded  as {\it primary}, i.e., implied by the form of the Lagrangian.

Motivated by devising an ideal
clock, Staruszkiewicz ob\-ser\-ved \cite{bib:astar2008} that (unlike unitarity) the irre\-du\-ci\-b\-i\-li\-ty of quantum systems has a
classical counterpart realized in postulating \eqref{eq:fundcond} as a means to singling out physically appealing Lagrangians. This postulate 
is in essence equivalent to the earlier, strong conservation idea due to Kuzenko, Lya\-kho\-vich and Segal \cite{bib:KLS1995}, introduced  
as a basic dynamical principle for devising Lagrangians suitable for geometric models of particles with spin. 

As established in \secref{sec:mech}, 
the phase space of the simplest clock 
can be parameterized using components of the position fourvector $x$ and  three tangent fourvectors $p,k,\Kpi$ bound to satisfy constraints \eqrefs{eq:fundcond2}{eq:fundcond}, where $$\SP{w}{w}\seq-\mathrm{Det}\left[
\begin{array}{@{}c@{\;}c@{\;}c@{}}
\SP{p}{p}&\SP{p}{k}&\SP{p}{\Kpi}\\
\SP{k}{p}&\SP{k}{k}&\SP{k}{\Kpi}\\
\SP{\Kpi}{p}&\SP{\Kpi}{k}&\SP{\Kpi}{\Kpi}
\end{array}\right]\weq \SP{p}{k}^2\SP{\Kpi}{\Kpi}.$$
Between these dynamical variables we assume a Poisson bracket $\PB{U}{V}\equiv\SP{\partial_{x}{U}}{\partial_{p}{V}}-\SP{\partial_{p}{U}}{\partial_{x}{V}}+\SP{\partial_{k}{U}}{\partial_{\Kpi}{V}}
-\SP{\partial_{\Kpi}{U}}{\partial_{k}{V}}$. 
\eqrefs{eq:fundcond2}{eq:fundcond} form a system of independent first class constraints with respect to this bracket. 
In line with Dirac method \cite{bib:dirac1950}, the most general Hamiltonian is 
a linear combination of all first class constraints with arbitrary functions $u$'s as coefficients. 
It is convenient that the combination be taken as:\footnote{
The original KLS Hamiltonian 
\cite{bib:KLS1995} involved a complex
variable  $\zeta$, ($\zeta\equiv \mcal{Z}(k)$), inherited from a primary Lagrangian. 
Starting with a related Lagrangian expressed in terms of $k$,
 a Hamiltonian analogous in form to \eqref{eq:hamiltonian} 
 was arrived at in \cite{bib:Ghosh} (upon earlier reducing an extended phase space).
Our approach goes in the opposite direction. We start with a Hamiltonian deduced  from first principles. In \cite{2012JPhCS.343a2017B} we generalized  this method onto systems described by a collection of fourvectors.}
\begin{multline}\label{eq:hamiltonian} \hspace{-0.05\linewidth}
H\seq\frac{u_1}{2\elm}\br{\SP{p}{p}-{\elm}^2} + \frac{u_2}{2\elm}
\br{\!\SP{p}{p}+\frac{4}{{\ell}^2{\elm}^2}\SP{k}{p}^2\SP{\Kpi}{\Kpi}}
\\
+u_3\SP{k}{\Kpi}+
        u_4\SP{k}{k}\end{multline}
The equations $\partial_{u_i}H\weq0$  form a system of first class constraints equivalent to \eqrefs{eq:fundcond2}{eq:fundcond}. Next, we introduce velocities $\dot{\wq}\weq\partial_{\wp}{H}$ \cite{bib:dirac1950}:
\begin{equation}\label{eq:DefVel}
\begin{split}
\hspace{-.08\linewidth} &\dx\weq\br{{u_1} + {u_2}}
  \frac{p}{\elm} +{u_2}\frac{4\,\SP{k}{p}\,\SP{{\Kpi}}{{\Kpi}}}
     {{{\ell}}^2\,{{\elm}}^3}\,k \weq {u_1} \frac{p}{\elm}+ {u_2}n,&\\
\hspace{-.08\linewidth}  & \dk\weq u_2\,\frac{4{\SP{k}{p}}^2}
   {{\ell}^2{\elm}^3}{\,\Kpi}+ u_3\,k,\hspace{.175\linewidth} n:={\frac{p}{\sqrt{\SP{p}{p}}}
  - \frac{\sqrt{\SP{p}{p}}\,k}{\SP{k}{p}}}.&
   \end{split}
   \end{equation}
By taking projective derivatives, defined recursively by 
$\udnp{n+1}{\slot}\eqdef\udp{(\udnp{n}{\slot})}$, where $\udp{\slot}\eqdef\br{\dot{\slot}}_{\perpp}$ and $\ts{(\slot)_{\perpp}\eqdef \slot-
\frac{\SP{p}{\slot}}{\SP{p}{p}}p}$, 
it can be shown that a curvature $\varkappa$ (defined by analogy with Frenet-Serret formulas) 
is fixed: $\varkappa\equiv- \SP{\udp{x}}{\udp{x}}^{-3}\gramm{\udp{x}}{\uddp{x}}\weq\sfrac{4}{\ell^2}$, 
 and that torsion vanishes  
on account of $\udp{x}$, $\uddp{x}$, and $\udddp{x}$ being coplanar as $p$-orthogonal linear combinations of
$p,k,\Kpi$. Hence, the trajectory perceived in the \cm frame 
is a circle of a fixed radius  $\sfrac{\ell}{2}$ (without constraints \eqref{eq:fundcond}, the radius would vary with the actual state \cite{bib:bratek2012H}). Correspondingly, the worldline's path is winding up 
around a fixed
space-time cylinder, the main axis of which represents the \cm inertial motion.
 To measure the rate of
change of the unit spatial vector $n$ in the \cm frame (see  \eqref{eq:DefVel} for the definition of  $n$), a frequency sca\-lar $\Omega$ can be introduced: \begin{equation}\label{eq:Omega}\Omega^2:={-\frac{\SP{\dot{n}}{\dot{n}}\SP{p}{p}}{\SP{p}{\dx}^2}};
\quad \Rightarrow\quad 
\Omega= \frac{\SP{p}{p}\sqrt{-\SP{\dk}{\dk}}}{\abs{
\SP{k}{p}\SP{p}{\dx}}}\;\; \mathrm{if}\;\; \dot{p}=0.\end{equation} For solutions, it reduces to a ratio:
$\Omega\weq\br{\sfrac{2}{\ell}}\abs{\sfrac{u_2}{u_1}}$, and if $\abs{u_1}>\abs{u_2}$ it is related
  to a hyperbolic angle $\Lambda$ between $p$ and $\dx$: $\Omega\weq\br{\sfrac{2}{\ell}}\tanh{\Lambda}$.
Both $\Omega$ and $\Lambda$ are re\-pa\-ra\-me\-tri\-za\-tion-invariant sca\-lars with 
obvious physical meaning. On the other hand, $\Omega$ and $\Lambda$ are functions of the arbitrary ratio
$\sfrac{u_2}{u_1}$. Thus the motion is indeterminate. 
To solve this paradox, this ratio needs to be set based on a sound guiding principle, so as not to
introduce arbitrary features into the dynamics.

\section{\label{sec:ranks}Singularities in the inverse Legendre transformation}

The form of a Lagrangian $L\equiv \SP{\dx}{p}+\SP{\dk}{\Kpi}-H$
corresponding to the Hamiltonian
\eqref{eq:hamiltonian}
is subject to invertibility of the
map \eqref{eq:DefVel}
restricted to a submanifold 
determined by the constraints \eqrefs{eq:fundcond}{eq:fundcond2}.
For the purpose of the invertibility analysis, it must suffice to focus upon Lorentz scalars only.
On the submanifold of interest, we may consider a map between two sets of variables:
$u_1$, $u_2$, $u_3$, $
\SP{k}{p}$, $\SP{p}{\Kpi}$ and 
$\SP{\dk}{\dk}$, $\SP{\dk}{\dx}$, $\SP{\dx}{\dx}$, $\SP{k}{\dx}$,
$\SP{k}{\dk}$:
%
\begin{equation}
\label{eq:map}
\begin{split}
&\SP{\dx}{\dx}\weq u_1^2-u_2^2,\qquad\qquad \SP{k}{\dx}\weq \frac{\SP{k}{p}}{\elm}\left(u_1+u_2\right),\\
&{{\SP{\dk}{\dx}}\weq  \ {
\frac{\SP{k}{p}}{\elm}   \left(
                \frac{4\,\SP{k}{p}\,\SP{p}{\Kpi}}{
{\elm}^3{\ell}^2}\,{u_2} +  {u_3} \right) }}\,\left( {u_1} + {u_2}
\right),\\
&\SP{\dk}{\dk}\weq -
\frac{4{\SP{k}{p}}^2}{{\ell}^2{\elm}^2}\,u_2^2,\quad\quad
\SP{\dk}{k}\weq0.
\end{split}
\end{equation}
The number of new constraints for velocities depends on the rank of the Jacobian matrix of this map.
Non-zero minors of maximal rank $4$ 
for this Jacobian are: {\scriptsize $j_1=\tfrac{16{\SP{k}{p}}^3}{{\ell}^2{\elm}^4}\, u_2
\br{u_1 + u_2}(u_2^2-u_1^2)$} and {\scriptsize $ j_2={\tfrac{4\,\SP{k}{p}}{{{\ell}}^2\,{{\elm}}^3}}{u_2}\,j_1$}.
Since $\SP{k}{p}\neq0$ (for a timelike $p$ and a null $k$), this implies that the Jacobian rank, $\mathrm{Rk}$, is dependent  on $u_{1,2}$. Full analysis distinguishes the following 4 regimes: 
\newcommand{\sep}{@{\hspace{2.5pt}}}
\begin{center}
\begin{tabular}{@{}|\sep c\sep|\sep c\sep|\sep c\sep|\sep c\sep|}
\hline
& Rk & & {velocity constraints}\\ \hline 
{\it i)} & $4$ &
  $u_1^2\neq u_2^2\ne0$ & $\SP{k}{\dk}\weq0$\\
{\it ii)} & $3$ & ${u_1}={u_2}\neq0$ & $\SP{k}{\dk}\weq0$,\ \ $\SP{\dx}{\dx}\weq0$,\ \ $ \ell^2\SP{\dk}{\dk}
+\SP{k}{\dx}^2\weq0$ \\
{\it iii)} & $2$ & ${u_1}=-{u_2}\neq0$ & $\SP{k}{\dk}\weq0$,\ \ $\dx\propto k$ $\quad \Rightarrow\quad \SP{\dx}{\dx}\weq0$\\
\hline
{\it ii')} & 3 & $u_2=0$,\  $u_1\ne0$ & $\SP{k}{\dk}\weq0$, \ \ $\SP{\dk}{\dk}\weq0$\\
\hline 
\end{tabular}
\end{center}
\smallskip\noindent
The  {\it ii'} case will not be of concern here, and $u_2\neq0$ is 
assumed   from now on. To find explicit expressions for momenta,  two cases are to be considered: $ u_1+u_2\neq0$ /{\it i, ii}/ or  $ u_1+u_2=0$ /{\it iii}/.

$\bullet$ For $u_2(u_1+u_2)\ne0$ 
the ansatz $p=\alpha_1 \dx+\alpha_2 k$ and $\Kpi=\beta_1 \dk +\beta_2 k$  allows to express momenta in terms of velocities and $u$'s. On substituting to \eqref{eq:DefVel} and solving for $\alpha_{1,2},\beta_{1,2}$, one gets:
\begin{align}\label{eq:momenta_gen}
\begin{split}
&p=\frac{{\elm}}{{u_1} + {u_2}}\,{\dot{x}} - 
   \frac{{{\ell}}^2\,{\elm}\,{\left( {u_1} + {u_2} \right) }^2\,
       \left( \SP{{\dot{k}}}{{\dot{k}}} - 2\,\SP{k}{\dk}\,{u_3} \right) }{4\,
       {\SP{k}{\dx}}^2\,{u_2}}\,\frac{k}{\SP{k}{\dx}},\\
&{\Kpi}= \frac{{{\ell}}^2\,{\elm}\,{\left( {u_1} + {u_2} \right) }^2}
     {4\,{\SP{k}{\dx}}^2\,{u_2}}\,\left( {\dot{k}} - k\,{u_3} \right).
\end{split}
\end{align}
The $\Psi_3$ constraint leads to $\SP{k}{\dk}\weq0$ (consistently with $\Psi_4$), while 
the $\Psi_{1,2}$  constraints give conditions for $u_{1,2}$: 
$$
\tfrac{1}{{\br{u_1+u_2}}^2} \SP{\dx}{\dx}+\tfrac{u_1+u_2}{2u_2}\,\xi=1
\quad\wedge\quad
 \tfrac{\br{u_1 + u_2}^2}{4\,{u_2}^2}\,\xi
       =1. \quad
       \boxed{\xi:=-\ell^2\tfrac{\SP{\dk}{\dk}}{{\SP{k}{\dx}}^2}}
$$
The resulting $u_{1,2}$ can be expressed as independent functions of  velocities, provided that the Jacobian determinant
${\scriptsize \tfrac{\partial(\Psi_1,\Psi_2)}{\partial(u_1,u_2)}}$, equal to $
{\scriptsize {\frac{-\elm^6\ell^2\,\xi
    }{4\,u_2^3\,\br{u_1 + u_2}}\, \SP{\dx}{\dx}}}$,
is nonzero, which leads to    
a {\it Lagrangian of the first kind}.
Otherwise, if $\SP{\dx}{\dx}=0$, 
then one gets $u_1=u_2$ and a frequency constraint $\SP{k}{\dx}^2+\ell^2\SP{\dk}{\dk}\weq0$.
This leads to a  {\it Lagrangian of the second kind}.

$\bullet$ For $u_2\neq0$ and $u_1+u_2=0$, one is led to a {\it Lagrangian of the third kind}
with $\dx\propto k$ (when $\SP{\dx}{\dx}\weq0$, $\SP{k}{\dx}\weq0$ and $\SP{\dk}{\dx}\weq0$). 

\subsection{\label{sec:nwp} Null worldlines principle}

Above, the rank of the inverse Legendre transformation, qualitatively dis\-cri\-mi\-nated between two separate regimes: $\SP{\dx}{\dx}\neq0$ (maximal rank) and $\SP{\dx}{\dx}=0$ (lower ranks).
Now, two other premises can be brought to the attention, as to why specifically the condition $\SP{\dx}{\dx}=0$ 
is so particular. 

In the maximal rank case, assuming any constraints 
such that  $\SP{\dx}{\dx}\neq0$ would be a matter of
arbitrary decision. For $\SP{\dx}{\dx}>0$, choosing a given function for $\Omega$ is equivalent to setting the hyperbolic angle $\Lambda$. But there is no privileged 
hyperbolic angle in the
(homogeneous) Lobachevsky space of fourvelocities (a similar argument on de Sitter hyperboloid  would apply to the 'tachionic' sector $u_2^2>u_1^2$). On the contrary, null worldlines  are
distinguished  by the lightcone stru\-cture of the spacetime. 
With
 $\SP{\dx}{\dx}=0$, the velocity can be fixed in a manifestly relativistically invariant manner independently of parameterization.  
We stress this important circumstance, since outside the light cone, a more general condition $\SP{\dx}{\dx}=\sigma$ with a given nonzero function $\sigma$, neither would set a velocity nor be reparameterization invariant. This qualitative difference should find its reflection also in the structure of the respective Lagrangians.\footnote{
This difference is already seen for  a material point
 described by  a general Lagrangian  $L=\frac{1}{2}({w^{-1}{\SP{\dx}{\dx}}+m^2 w})$. The equation $\partial_wL=0$ implies two qualitatively distinct regimes: 1) in which $w$ is a function of $\dx$, then $w=m^{-1}\sqrt{\SP{\dx}{\dx}}$,
and 2) in which $w$ is independent of $\dx$, requiring $m=0$ and  
a constraint $\SP{\dx}{\dx}=0$. The resulting Lagrangians are:
1) that of a massive particle $L=m\sqrt{\SP{\dx}{\dx}}$ with $p=m\frac{\dx}{\sqrt{\SP{\dx}{\dx}}}$  and 2) that of a massless particle $L=\frac{1}{2}w^{-1}\SP{\dx}{\dx}$ with $p=w^{-1}\dx$ and an arbitrary $w$ transforming as $\delta{w}=w\delta \epsilon$ under a reparameterization $\delta{\dx}=\dx \delta\epsilon$. The analytic form of $L=m\sqrt{\SP{\dx}{\dx}}$ would not be suitable in a region containing the surface $\SP{\dx}{\dx}=0$, where the corresponding $p$ would be divergent.   }
Yet, there is an insightful  remark due to Dirac, showing the distinguished role of the condition $\SP{\dx}{\dx}=0$. It is a
[counterintuitive] consequence of Dirac equation, that {\it a measurement of the electron's
instantaneous motion 
is  certain to give the speed of light},
which Dirac mentions in his {\it Principles} 
\cite{bib:dirac:principles} 
and asserts this result to 
be generally true
in a relativistic theory. 

The Dirac observation in conjunction with previous findings tempts one
to conjecture
that 
{\it worldlines of classical analogs of quantum elementary particles should be~null}.

\section{\label{sec:maxreg}\label{sec:FRR} 
Lagrangians of the first kind}

In the {\it sub-luminal} sector ($u_1^2>u_2^2$) let 
$u_1=\rho\,\cosh{\psi}$,
$u_2=\rho\,\sinh{\psi}$, $\rho>0$, $\abs{\psi}<\infty$.  Then from
\eqref{eq:map}: 
${\scriptstyle\rho=\etasign\sqrt{\SP{\dx}{\dx}}}$, 
${\scriptstyle\tanh{\psi}=\mp\, \frac{\sqrt{\xi }}{2 {\pm}{\sqrt{\xi }}}}$. 
With the resulting $u_{1,2}$ substituted in \eqref{eq:momenta_gen},  two Lagrangians follow:
 $\tilde{L}=\etasign\elm
\sqrt{\SP{\dx}{\dx}} \sqrt{ 1\pm\sqrt{\xi}
}+\lambda_1\SP{k}{k}+\lambda_2\SP{k}{\dk}$ ($\lambda_{1,2}$ involve arbitrary  $u_{3,4}$).

 In the {\it super-luminal} sector ($u_1^2<u_2^2$)  -- which may be considered on account of $x$ not being  assigned to a \cm motion -- 
a similar analysis (with $u_1=-\etasign\hat{\rho}\,\sinh{\hat{\psi}}$ and
$u_2=-\etasign\hat{\rho}\,\cosh{\hat{\psi}}$, $\hat{\rho}\neq0$)
leads to a single Lagrangian 
$\tilde{L}=\etasign\elm
\sqrt{-\SP{\dx}{\dx}} \sqrt{\sqrt{\xi}-1
}+\lambda_1\SP{k}{k}+\lambda_2\SP{k}{\dk}$.

In both cases, the last term in $\tilde{L}$ (whose only effect is an  additive gauge-like term
in the canonical momentum
$\partial_{\dk}\tilde{L}\to\partial_{\dk}\tilde{L}+\alpha k$)
can be integrated
off by parts. On denoting the 
remaining term $(\lambda_1-\br{\sfrac{1}{2}}\dot{\lambda}_2)\SP{k}{k}$ by $ \lambda\SP{k}{k}$,  
one finally ends up with two Lagrangians (equivalent to those arrived at in \cite{bib:KLS1995,bib:astar2008}):
\begin{equation}\label{eq:LagrFRR}L_{\pm}=
\etasign\elm \sqrt{\SP{\dx}{\dx}\br{ 1\pm\sqrt{- {\ell}^2\frac{
\SP{\dk}{\dk} }{ {\SP{k}{\dx}}^2 } }}\; }+\lambda\,
\SP{k}{k},\end{equation}
with their respective 
Lagrange multipliers $\lambda$.
The sub-luminal  Lagrangian $L_{+}$ is that 
of the Fundamental Relativistic Rotator \cite{bib:astar2008}. With the Lagrangian
$L_{-}$ we could consider both sub- or super-luminal motions.

In the clock context, it  is appropriate to recall an earlier result \cite{2009arXiv0902.4189B} published in \cite{bib:bratek2011} that the  Lagrangians \eqref{eq:LagrFRR}  can be alternatively  arrived at by adopting a physically dubious condition 
that the Hessian matrix ${\partial}_{\wdq\wdq}L$ for a general Lagrangian $L=f(\xi)\sqrt{\SP{\dx}{\dx}}$ expressed in terms of  only the $5$ degrees of freedom 
characteristic of a rotator -- Cartesian $\vec{x}(t)$ and spherical $\vartheta(t),\varphi(t)$  (considered as functions of $x^0\equiv t$) -- must be zero. 
\newcommand{\dc}{f_{{}^{,\xi}}}
\newcommand{\ddc}{f_{{}^{,\xi\xi}}}
As shown therein, this leads to a differential equation for $f$: $\dc f+2\xi({\dc}^2+\ddc f)=0$.
As a direct consequence of this, the clocking frequency becomes indeterminate. This conforms with what has been concluded in \secref{sec:hamiltonian}.  
For reasons described in \secref{sec:nwp}, with  the Lagrangian \eqref{eq:LagrFRR}, there would be no privileged velocity constraint suitable to set this frequency so as to make the motion determinate, while conditions involving $\SP{\dx}{\dx}=0$ (such as {\it ii} or {\it iii}) would not be compatible with the analytic structure of these Lagrangians  (the canonical momenta  $\partial_{\wdq}L$ would involve indeterminate forms $0/0$). For these reasons we must come to the conclusion that \eqref{eq:LagrFRR} does not describe a clock at all.  

  It seems that neither considering more complicated systems \cite{Bratek:2009wc,2012JPhCS.343a2017B} nor 
introducing interactions \cite{bib:bratek2011} would help to remove this indeterminacy of motion. For example, in the electromagnetic field, the consistency requirements $\PB{\Psi_{1,2}}{H}\weq0$  (with $p-eA$ substituted for $p$ by the minimal coupling principle) lead to a secondary constraint $F_{\mu\nu}p^{\mu}k^{\nu}\weq0$, which for rotators reduces  to  a condition $F_{\mu\nu}\dx^{\mu}k^{\nu}=0$  strictly connected with the Hessian singularity 
alluded to above. Although this condition might lead to a unique motion in some situations (e.g. with appropriate initial data in a uniform magnetic field \cite{bib:kassandrov2009}) this may not to be so in general (see, a toy model \cite{bib:bratek2010toy}).

\section{Ideal Clocks}

\subsection{\label{bib:secLag}Second kind Lagrangian}

The new velocity constraints arranged  to forms of the first degree in the velocities read:
\begin{equation}\label{eq:cnstrI}
\frac{\SP{\dx}{\dx}}{\SP{k}{\dx}}\weq0,\quad
 \ell^2 \frac{\SP{\dk}{\dk}}
    {\SP{k}{\dx}} + \SP{k}{\dx}\weq0.
\end{equation}
By eliminating these constraints from \eqref{eq:map}, one finds
$u_1=\chi$, $u_2=\chi$, $u_3=\upsilon$, 
 $\scriptstyle {\SP{k}{p}=
\frac{\elm\SP{\dot{x}}{k}}{2\,\chi}}$ and 
$\scriptstyle \SP{p}{{\Kpi}} = 
   \frac{{{\ell}}^2\,{{\elm}}^2}{2\,\SP{k}{\dx}}\,
    \left( \frac{\SP{{\dot{k}}}{{\dot{x}}}}{\SP{k}{\dx}} - \upsilon  \right) $, where
$\chi$ and $\upsilon$ are arbitrary functions.
After discarding a total derivative involving $\SP{k}{\dk}$ and 
the higher order terms in the velocity constraints (irrelevant for the Dirac variational procedure \cite{bib:dirac1950}), the resulting Lagrangian  
can be arranged in a form with a new independent variable $\scriptstyle \kappa(\chi)\equiv \frac{\SP{k}{p}}{\elm}$ and a Lagrange multiplier $\lambda$:
\begin{equation}\label{eq:LagI}
\boxed{L=\frac{\elm\kappa}{2} \frac{\SP{\dx}{\dx}}{\SP{k}{\dx}} +
\frac{\elm}{4 \kappa}\left( \ell^2\frac{
\SP{\dk}{\dk}}{\SP{k}{\dx}} + \SP{k}{\dx} \right)   + \lambda\, \SP{k}{k}.}
\end{equation}
As expected,  this Lagrangian is linear in the velocity constraints, with functions of momenta as coefficients.
In view of the equation $\partial_{\kappa}L=0$, the conditions
\eqref{eq:cnstrI} can be regarded as consequences of one another, and hence, only $\SP{\dx}{\dx}=0$ may be imposed as a subsidiary condition. Then $\kappa$ becomes arbitrary. Conversely, if $\partial_{\kappa}L=0$ is to be satisfied for arbitrary $\kappa$, then both conditions in \eqref{eq:cnstrI} follow.
The Ca\-si\-mir invariants 
 ${\scriptstyle \SP{p}{p}=\frac{\elm^2}{2}\br{1+\xi}}$
 and 
${\scriptstyle \SP{w}{w}=-\frac{\ell^2\elm^4}{4}\xi}$ 
 are bound to satisfy only a single constraint
${\scriptstyle \SP{p}{p}\weq\frac{\elm^2}{2}-\frac{2}{\ell^2\elm^2}\SP{w}{w}}$
and off the surface ${\scriptstyle \frac{{{\ell}}^2\,\SP{\dk}{\dk}}
    {\SP{k}{\dx}} + \SP{k}{\dx}\weq0}$ they
would be functions of the velocities. But for \eqref{eq:LagI} the principal conditions are satisfied on the basis of Hamilton's principle, either supplemented with the null worldlines principle or with the condition that $\kappa$ be independent of the velocities.\footnote{Because $\SP{k}{\partial_{\dx}L}\equiv m\kappa$,  
the freedom in choosing $\kappa$ at an instant  (with $k$ being set) involves the freedom in choosing a combination of momentum variables in $p$. This dependence of a Lagrangian on momentum variables is characteristic of systems with velocity constraints \cite{bib:dirac1950}.}
The latter requirement is crucial, since otherwise, by solving $\partial_{\kappa}L=0$ for $\kappa$, one would end up with a qualitatively different Lagrangian ${\scriptstyle \elm\sqrt{\frac{\SP{\dx}{\dx}}  {2}\br{1+\ell^2\frac{\SP{\dk}{\dk}}{\SP{k}{\dx}^2}}}\,+\lambda\SP{k}{k}}$ whose analytic form is not admissible
 on the surface \eqref{eq:cnstrI} (the momenta  $\partial_{\wdq}L$ would involve indeterminate forms $\frac{0}{0}$).   
  
\subsubsection{Connection with a family of Relativistic Rotators.}
To extend  the construction in \cite{bib:astar2008} so as to include also the case of   \secref{bib:secLag}, let
 a class of projection invariant La\-g\-ran\-gians of the first degree in the velocities be considered, whose form   would be admissible also on the surface  $\SP{\dx}{\dx}=0$ and compatible with the condition $\SP{k}{\dx}\neq0$: 
\begin{equation}\label{eq:rotators}L_{\mcal{F}}=\frac{\elm{\kappa}}{2}\,\frac{\SP{{\dx}}{{\dx}}}{\SP{k}{\dx}}
+\frac{\elm}{2{\kappa}}\SP{k}{\dx} \mcal{F}(\xi) + \lambda\, \SP{k}{k}.\end{equation}
The ${\kappa}$ must have appeared in this precise way for the dimensional grounds and it must transform as $\kappa\to \alpha \kappa$ when $k\to\alpha k$ on account of the assumed projection invariance. Here, $\mcal{F}$ is any function.
If $\partial_{{\kappa}}{L_{\mcal{F}}}=0$,  the principal constraints reduce to $\mcal{F}(\xi ) - 2\,\xi \,\mcal{F}'(\xi )=1$
   and $4\,\xi \,{\mcal{F}'(\xi )}^2=1$ for any $\kappa$.
If $\kappa$ is not a function of velocities, then $\partial_{{\kappa}}{L_{\mcal{F}}}=0$ implies $\mcal{F}=0$ (and $\SP{\dx}{\dx}=0$), 
then the principal conditions give $\mcal{F}'=-\frac{1}{2}$ and $\xi=1$. Hence, to a linear order, $\mcal{F}(\xi)=
\br{1-\xi}/2+o(1-\xi)$ in the vicinity of $\xi=1$. And this is another way of arriving at \eqref{eq:LagI}. 
In contrast, for ${\kappa}$ not independent of the velocities, one would conclude from $\partial_{{\kappa}}L_{\mcal{F}}=0$ that ${\kappa}={\SP{k}{\dx}}{}\sqrt{\mcal{F}(\xi)/\SP{\dx}{\dx}}$ and end up with a class of Lagrangians  $\elm
   {\sqrt{\SP{{\dot{x}}}{{\dot{x}}}\mcal{F}(\xi)}}$ 
describing relativistic rotators considered in \cite{bib:astar2008} (which includes
Lagrangians $L_{\pm}$ of \secref{sec:maxreg} as a special case with $\mcal{F}(\xi)=1\pm\sqrt{\xi}$). 

\subsection{\label{sec:3rdclass}  Third kind Lagrangian}

Putting $u_1=-u_2$, consider for a while a restricted Legendre transformation with 
$p$ left unaltered. Taking ${\scriptstyle
\Kpi=\mp\frac{\ell\elm^2}{2}
     \frac{\dk-k u_3}{\SP{k}{p}\sqrt{-\SP{\dk}{\dk}}}}$ and ${\scriptstyle u_2=\mp
     \frac{\ell\elm}{2 \SP{k}{p}}\sqrt{-\SP{\dk}{\dk}}}$ implied by \eqrefs{eq:DefVel}{eq:map} into account (where ${\scriptstyle \sgn{\frac{\SP{p}{\dx}}{\SP{k}{p}}}=\pm1}$) and integrating off the term linear in  $\SP{k}{\dk}$, one arrives at
  a Lagrangian:     \begin{equation}\label{eq:lagB1}
L= \SP{\dx}{p} \pm\,
  \frac{\ell \elm^2}{2}  \frac{ 
    \sqrt{-\SP{\dk}{\dk}}}{ \SP{k}{p}} 
+\lambda\SP{k}{k}.\end{equation}
By making arbitrary variations w.r.t. $p$ ($\delta L$ must be independent of $\delta p$ \cite{bib:dirac1950}),
 the result ${\scriptstyle 
     \dx = \pm\frac{\ell\elm^2}{2}
\frac{\sqrt{-\SP{\dk}{\dk}}}{\SP{k}{p}^2}k}$ following from \eqrefs{eq:DefVel}{eq:map} can be re-obtained.
It implies $\scriptstyle{\SP{{\dot{x}}}{e}} =\pm
   \frac{{\ell}\,{{\elm}}^2\,{\sqrt{-\SP{{\dot{k}}}{{\dot{k}}}}}}
      {2\,{\SP{k}{p}}^2}\SP{e}{k}$ for any
vector $e$, and this fact can be used to eliminate $p$ from $L$. 
Hence, 
the alternative Lagrangian takes
                    on a form involving arbitrary $e$ such that $\SP{k}{e}\neq0$:
\begin{equation}\label{eq:lagB2}
L={\elm}\,{\etasign}\,
   {{\sqrt[4]{-{\frac{\,4\,{\ell}^2\,\SP{{\dot{x}}}{e}^2\,{\SP{{\dot{k}}}{{\dot{k}}}}}
       {\SP{e}{k}^2}}}}}
+\lambda \SP{k}{k}.\end{equation}
For 
$\SP{e}{k}$ to be nonzero, it would 
suffice that $e$ be obtained from any timelike
vector by a two-parameter transformation group $e\to \alpha
(e+\beta k)$,  with $\alpha$, $\beta$ being arbitrary functions.
This freedom in choosing 
$e$ must be 
physically irrelevant, and this will be so if $\partial_e L=0$. 
This
implies $\dx=\frac{\SP{\dx}{e}}{\SP{k}{e}}k$.
Furthermore, $p:= \partial_{\dx}L$ is collinear with $e$ and is independent of the scale of $e$.
As so, $p$ may be substituted  in place of $e$ in  the expression for $\partial_{\dx}L$, hence \begin{equation}\label{eq:freq}\frac{2}{\ell} =
   \frac{{{\elm}}^2\,{\sqrt{-\SP{{\dot{k}}}{{\dot{k}}}}}}{\abs{\SP{{\dot{x}}}{p}\SP{k}{p}}} \quad\Rightarrow\quad \Omega=\frac{2}{\ell}\frac{\SP{p}{p}}{\elm^2} \quad \mathrm{(from\, \eqref{eq:Omega})}.\end{equation}
 The 
constraint $\SP{p}{p}-\elm^2\weq0$ does not follow from  the Lag\-ran\-gians \eqrefs{eq:lagB1}{eq:lagB2}, nevertheless it is essential for consistency with the map \eqref{eq:map}. It must be regarded as a {\it secondary} first class constraint (whose purpose is to set $\Omega$ to
$\sfrac{2}{\ell}$ and the orbital radius to $\sfrac{\ell}{2}$, consistently with the equations of motion).

\subsection{Comparison of the clocks} It is convenient to write down the Hamiltonian equations in the \cm gauge: $\SP{p}{\Kpi}=0$,
$\SP{k}{p}=\elm$, $\SP{p}{\dx}=\elm$ and to consider a unit space-like direction $n$ defined in \eqref{eq:DefVel}, which is collinear with the projection of $k$ onto a subspace orthogonal to $p$. Together with the consistency requirements
$0=\PB{\SP{k}{p}}{H}$
and
$0=\PB{\SP{p}{\Kpi}}{H}$, this 
implies for $u_1=\pm u_2$ that $u_1=1$, $u_2=\pm 1$, $u_3=0$ and $u_4=
   \pm\frac{\elm }{2}$. 
 This way the
Hamiltonian equations reduce to
$$ \dx=\frac{p}{\elm}\mp n , \qquad
\dot{p}=0,\qquad \dot{n}=\pm\frac{4}{\elm\,\ell^2}\,\Kpi,\qquad
\dot{\Kpi}=\mp\elm\,n,$$  
with $\SP{n}{n}=-1$, $\SP{n}{p}=0$ (then $k=\frac{p}{\elm}+n$). 
The equations for $\dot{n}$ and $\dot{\Kpi}$ imply a
uniform motion with frequency $\Omega=\frac{2}{\ell}$ about a great circle on the unit sphere:  $\ddot{n}+\frac{4}{\ell^2}\,n =0$. 
The null directions of the clocks' velocity vectors $\dx$ are conjugate to one another by the reflection $\frac{\elm\dx}{\SP{p}{\dx}}\to\frac{2p}{\elm}-\frac{\elm\dx}{\SP{p}{\dx}}$. 
Interestingly, the two clocks 
have opposite spin alignment:
$$p\wedge k\wedge\Kpi=\pm\frac{\elm\ell^2}{4}p\wedge n\wedge{\dot{n}}=\pm\elm\,p\wedge x\wedge{\dot{x}},$$
where $x=\frac{p}{\elm}t+\frac{\ell}{2}r(\varphi)$, $\SP{r}{r}=-1$, $\SP{p}{r}=0$, $\varphi=\frac{2}{\ell}t$ (then $n=\pm r'(\varphi)$). 
In a sense, the two clocks can be regarded as a limiting case of the Lagrangian   \eqref{eq:LagrFRR} when
$\SP{\dx}{\dx}\to0$ with: {\it a)} $\frac{\ell^2\SP{\dk}{\dk}}{\SP{k}{\dx}^2}\to -1$ for the clock \eqref{eq:LagI} or {\it b)} $\SP{k}{\dx}\to0$ for the clock \eqref{eq:lagB2}.

\section{Summary and future applications}

In this paper were described two mathematical clocks which are relativistic rotators  exhibiting intrinsic circular motion with the speed of light and opposite spin alignment.  The Lagrangians of the clocks were distinguished by a singularity of an inverse Legendre map for rotators of which Casimir scalars are fixed parameters. Such clocks are perfect, they work always the same way, no matter how they move. 

In future works, the two ideal clocks can be used to test the clock hypothesis.
In free motion, the phase associated with the intrinsic circular motion of these clocks increases stea\-di\-ly with the affine parameter of the center of mass (\cm). But it is not a priori obvious  (even in the limit $\ell\to0$) if this property will survive for accelerated motions of the \cm, e.g. for a constrained motion along a strongly curved worldline. 
For such  motions, the {\it chronometric curve} -- that is, properly parameterized helical null worldline of an ideal clock -- would undergo additional distortions and this could affect the steady clocking rate. 

Testing the clock hypothesis requires introducing interactions.  However, usual coupling with external fields may lead to inconsistencies, see \cite{Dirac:1971:PERb}. In this context, it would be instructive to see  the implications of the secondary constraint $F_{\mu\nu}p^{\mu}k^{\nu}\weq0$ appearing when ideal clocks are minimally coupled with the electromagnetic field.  Furthermore,
it would be interesting  to study the motion in strong gravitational fields of black holes. 
In curved spacetimes, there might arise problems even with defining the rotation phase: when
global teleparallelism is lost, the local reference frames, used to measure the infinitesimal phase increments at various instants, cannot  be unambiguously connected; in addition, some disturbances in the phase could appear due to rotation of local inertial frames.

\bibliographystyle{ieeetr}\bibliography{IdealClock}

\end{document}